\documentclass[12pt]{article}
\usepackage{amsthm} 
\usepackage{amsmath} 
\usepackage{amsfonts}
\usepackage{amsxtra}
\usepackage{yhmath}
\usepackage{amssymb} 
\usepackage{mathrsfs} 
\usepackage{esint} 
\usepackage{empheq}
\usepackage{stmaryrd}
\usepackage{dsfont}
\usepackage[latin1]{inputenc}
\usepackage{upgreek}
\usepackage[all]{xy}
\usepackage{graphicx}
\usepackage{latexsym}
\usepackage{color}
\usepackage{epsfig}
\usepackage{graphicx}
\usepackage{caption}
\usepackage{subcaption}
\usepackage{here}
\usepackage{float}
\usepackage{array}
\usepackage{tikz}
\usepackage{tikz-cd}
\usetikzlibrary{3d,arrows,calc,chains,matrix,positioning,scopes}

\DeclareSymbolFont{UPM}{U}{eur}{m}{n}
\DeclareMathSymbol{\uppartial}{0}{UPM}{"40}

\def\ba{\begin{array}}
	\def\ea{\end{array}}

\def\bea{\begin{eqnarray}}
	\def\eea{\end{eqnarray}}

\newcommand{\Hom}[2]{\mathrm{Hom}\mathopen{}\left(#1,#2\right)}

\theoremstyle{definition}

\theoremstyle{plain}

\begin{document}
	
	\pagestyle{empty}
	\setcounter{page}{0}
	\hspace{-1cm}

	\begin{center}
		{\Large {\bf The U(1) BF functional measure and the Dirac distribution on the space of quantum fields}}%
		\\[1.5cm]
		
		{\large F. Thuillier}
		\vspace{3mm}
		
		{\it Université Grenoble Alpes, USMB, CNRS, LAPTh, F-74000 Annecy, France, frank.thuillier@lapth.cnrs.fr}
	\end{center}
	
	\vskip 0.7 truecm

	\vspace{3cm}
	
	\centerline{{\bf Abstract}}
	In this letter, we explain how the U(1) BF measure can be related to the Fourier transform of a Dirac distribution defined on the $\mathbb{Z}$-module of quantum fields. Then, we revisit the U(1) BF partition function with the help of this Dirac distribution and finally shed light on a natural relation between the U(1) BF and Chern-Simons theories.
	
	\vspace{2cm}
	
	\vfill
	
	\newpage
	\pagestyle{plain} \renewcommand{\thefootnote}{\arabic{footnote}}
	
	\section{Introduction}
	
	The study of the $U(1)$ BF and Chern-Simons theories in the context of Deligne-Beilinson cohomology has been extensively developped in a long series of articles \cite{GT1,GT2,GT3,MT}. A rephrasing as well as a dimensional extension of the $U(1)$ BF theory have been proposed in \cite{HMT22}. In this later context, Deligne-Belinson cohomology classes are called quantum fields and the quantum lagrangian of the theory is the Deligne star product of a quantum $p$-field with a quantum $q$-field, with $p+q+1=n$ the dimension of the closed smooth oriented manifold $M$ on which the Lagrangian is considered. Unlike a classical Lagrangian, this quantum Lagrangian is defined modulo forms with integral periods, hence the qualifier quantum, the corresponding quantum action thus being defined modulo integers. The partition function of the $U(1)$ BF theory on $M$ is then
	\begin{equation}
		\label{kpartition}
		\mathcal{Z}_{BF_k} = \frac{1}{\mathcal{N}} \iint \! \! D\!A \,  D\!B \, e^{2 i \pi k \oint_M A \star B} \, ,
	\end{equation}
	where $k \in \mathbb{Z}$ is the coupling constant of the theory. Of course, a precise meaning of the measure $D\!A$ has to be given and a precise choice for the normalization factor $\mathcal{N}$ has to be done. Remarkably, thanks to the exact sequences into which the space of quantum fields sits, it is possible to write $D\!A$ as a product the first factor of which is a finite dimensional measure while the second, although remaining purely infinite dimensional, is very close to the functional measure met in QFT. Then, by choosing $\mathcal{N}$ to be the contribution to the functional integral produced by this second factor, we are left with a determination of $\mathcal{Z}_{BF_k}$ which relies on well-defined integrals and sums. In the $3$-dimensional case, this yields
	\begin{equation}
		\mathcal{Z}_{BF_k} = |T_1(M)| |H^1(M,\mathbb{Z}_k)| \, ,
	\end{equation}
	where $|T_1(M)|$ is the order of the torsion group $T_1(M)$ and $|H^1(M,\mathbb{Z}_k)|$ the order of the group $H^1(M,\mathbb{Z}_k) = \Hom{H_1(M)}{\mathbb{Z}_k}$. If we use the standard decomposition $T_1(M) = \mathbb{Z}_{p_1} \oplus \mathbb{Z}_{p_2} \oplus \dots \oplus \mathbb{Z}_{p_N}$, with $p_i$ dividing $p_{i+1}$, then we have \cite{MT}
	\begin{equation}
		\mathcal{Z}_{BF_k} = \prod_{i=1}^N p_i \gcd(k,p_i) \, .
	\end{equation}
	A similar approach can be made to determine the expectation values of the quantum observables of the theory. However, a different normalization factor is used in that case. With this choice of normalization, the expectation value of the trivial observable yields the partition function itself instead of $1$ \cite{MT2}. Let us point that a similar change of normalization appears in the $3$-dimensional $U(1)$ Chern-Simons theory \cite{GT2,GT3}. 
	
	Interestingly, when $k=1$ we have
	\begin{equation}
		\mathcal{Z}_{BF_k} = |T_1(M)| \, ,
	\end{equation}
	since in that case $\mathbb{Z}_1 = \{1\}$, the only element of $\Hom{H_1(M)}{\mathbb{Z}_k}$ thus being the trivial homomorphism. The above result can also be written as
	\begin{equation}
		\frac{1}{|T_1(M)|}\mathcal{Z}_{BF_k} = 1 \, ,
	\end{equation}
	and if we consider expression \eqref{kpartition} of the partition function, it seems natural to wonder whether we can write
	\begin{equation}
		\label{deltaonfields}
		\int \! \! D\!A \, e^{2 i \pi \oint_M A \star B} = \delta(B) \, ,
	\end{equation}
	with $\delta(B)$ the Dirac distribution on the space of quantum fields, just like the usual Dirac distribution (centered at the origin in $\mathbb{R}^n$) is the Fourier transform of the constant function $1$. This article tries to answer this question.
	
	Let us end this introduction by recalling that the choice to define the quantum Lagrangian as a Deligne star product of quantum fields automatically implies that the coupling constant $k$ of the $U(1)$ BF and Chen-Simons theories is quantized, i.e., that $k \in \mathbb{Z}$.
	
	All along this letter $M$ is a smooth oriented closed $n$-dimensional manifold.

	\section{Abelian BF theory}
	
	As already mentioned in the introduction, the quantum Lagrangian of the $U(1)$ BF theory is defined as the Deligne product of two quantum fields: $A \star B$. The corresponding quantum action is then
	\begin{equation}
		S[A,B] = \oint_M A \star B \, ,
	\end{equation}
	which is $\mathbb{R}/\mathbb{Z}$-valued. We won't go into the detailed construction of quantum fields, referring the reader to the standard mathematical literature on Deligne-Beilinson cohomology or to \cite{HMT22} if the reader wants to stick with the quantum fields terminology\footnote{A gauge field is a $n$-tuple whose entries are collections of local forms fulfilling a set of descent equations. Gauge field transformations form a subset of the set of gauge fields. By identifying gauge fields which differ by gauge field transformations we obtain quantum fields.}. What we need in this letter is the following exact sequence in which the space of quantum $p$-fields, $H_D^p(M)$, sits
	\begin{equation}
		\label{short1}
		0 \rightarrow \frac{\Omega^p(M)}{\Omega_\mathbb{Z}^p(M)} \rightarrow H_D^p(M) \rightarrow H^{p+1}(M) \rightarrow 0 \, ,
	\end{equation} 
	where $\Omega^p(M)$ denotes the space of $p$-forms of $M$, $\Omega_\mathbb{Z}^p(M)$ the subspace of closed $p$-forms with integral periods and $H^{p+1}(M) = F^{p+1}(M) \oplus T^{p+1}(M)$ is the $(p+1)$-th cohomology space of $M$. The above exact sequence implies that we can identify, in a non-canonical way, any quantum $p$-field with a triplet $(n_A , \tau_A , \hat{\omega}_A)$ where $n_A \in F^{p+1}(M)$, $\tau_A \in T^{p+1}(M)$ and $\hat{\omega}_A \in \Omega^p(M)/\Omega_\mathbb{Z}^p(M)$. It is possible to go one step further by decomposing, still in a non-canonical way, each $\hat{\omega}_A$ according to the exact sequence
	\begin{equation}
		\label{short2}
		0 \to \frac{\Omega_\circ^p(M)}{\Omega_\mathbb{Z}^p(M)} \rightarrow \frac{\Omega^p(M)}{\Omega_\mathbb{Z}^p(M)} \xrightarrow{\pi} \frac{\Omega^p(M)}{\Omega_\circ^p(M)} \to 0 \, ,
	\end{equation}
	where $\Omega_\circ^p(M)$ denotes the space of closed $p$-forms of $M$. This means that we can associate with $\hat{\omega}_A$ a couple $( \theta_A , \bar{\omega}_A)$ with $\theta_A \in \Omega_\circ^p(M)/\Omega_\mathbb{Z}^p(M) = (\mathbb{R}/\mathbb{Z})^{b_q}$, $q=n-p-1$ and $b_q$ being the $q$-th Betti number of $M$, and with $\bar{\omega}_A \in \Omega^p(M)/\Omega_\circ^p(M)$\footnote{More precisely, exact sequence\eqref{short1} describes $\Omega^p(M)/\Omega_\mathbb{Z}^p(M)$ as a fibre bundle over of $\Omega^p(M)/\Omega_\circ^p(M)$. So, if $\sigma$ is a continuous section of this bundle then for any $\bar{\omega} \in \Omega^p(M)/\Omega_\circ^p(M)$ we have $\sigma(\bar{\omega}) \in \Omega^p(M)/\Omega_\mathbb{Z}^p(M)$. Moreover, if $\hat{\omega}$ is an element of $\Omega^p(M)/\Omega_\mathbb{Z}^p(M)$, then $\theta = \hat{\omega} - \sigma \circ \pi(\hat{\omega}) = \hat{\omega} - \sigma (\bar{\omega})$ is an element of $\Omega_\circ^p(M)/\Omega_\mathbb{Z}^p(M)$.}. All in all, we have
	\begin{equation}
		\label{Adecomps}
		A \rightarrow (n_A , \tau_A , \hat{\omega}_A) \rightarrow (n_A , \tau_A , \theta_A , \bar{\omega}_A) \, .
	\end{equation}
	Once more, all the details can be found in \cite{HMT22}. We refer to the above $4$-tuple as the complete decomposition of $A$.
	
	According to correspondence \eqref{Adecomps}, the quantum Lagrangian $A \star B$ gives rise to the following contributions \cite{MT}
	\begin{align}
		&n_A \star n_B + n_A \star \tau_B + n_A \star \hat{\omega}_B + \nonumber \\
		+ &\tau_A \star n_B + \tau_A \star \tau_B + \tau_A \star \hat{\omega}_B +  \nonumber \\
		+ &\theta_A \star n_B + \theta_A \star \tau_B + \theta_A \star \hat{\omega}_B +  \nonumber \\
		+ &\bar{\omega}_A \star n_B + \bar{\omega}_A \star \tau_B + \bar{\omega}_A \star \hat{\omega}_B \, , \nonumber 
	\end{align}
	where, for now, we did not decompose $B$ completely. In fact, we have the following set of properties \cite{MT}
	\begin{equation}
		\oint_M n_A \star n_B = 0 \quad , \quad  \oint_M \tau_A \star \hat{\omega}_B = 0 \quad , \quad \oint_M \theta_A \star \hat{\omega}_B = 0 \, ,
	\end{equation}
	all these integrals taking their values in $\mathbb{R}/\mathbb{Z}$. Thus, the only non trivial contributions to consider are
	\begin{align}
		&n_A \star \tau_B + n_A \star \hat{\omega}_B + \nonumber \\
		+ &\tau_A \star n_B + \tau_A \star \tau_B +  \nonumber \\
		+ &\theta_A \star n_B + \nonumber \\
		+ &\bar{\omega}_A \star n_B + \bar{\omega}_A \star \hat{\omega}_B \, , \nonumber 
	\end{align}
	or if we completely decompose $B$
	\begin{align}
		&n_A \star \tau_B + n_A \star \theta_B + n_A \star \bar{\omega}_B + \nonumber \\
		+ &\tau_A \star n_B + \tau_A \star \tau_B +  \nonumber \\
		+ &\theta_A \star n_B + \nonumber \\
		+ &\bar{\omega}_A \star n_B + \bar{\omega}_A \star \bar{\omega}_B \, , \nonumber 
	\end{align}
	
	Associated with decomposition \eqref{Adecomps} of quantum fields, we have the natural measures
	\begin{equation}
		\sum_{n_A} \quad , \quad \sum_{\tau_A} \quad , \quad \oint d \theta_A \quad , \quad \int D \bar{\omega}_A \quad . 
	\end{equation}
	The first sum is performed on $F^{p+1}(M) \simeq F_q(M)$, where $q = n-p-1$, the second is over $T^{p+1}(M) = \mathbb{Z}_{p_1} \oplus \mathbb{Z}_{p_2} \oplus \dots \oplus \mathbb{Z}_{p_N} \simeq T_q(M)$, the third measure is over the torus $(\mathbb{R}/\mathbb{Z})^{b_q}$ and the final measure is over the infinite dimensional quotient $\Omega^p(M)/\Omega_\circ^p(M)$. This last measure is the only one to be a priori ill-defined. A work around to this problem is to consider a finite dimensional subspace of $\Omega^p(M)/\Omega_\circ^p(M)$, ``big" enough for our purpose, on which $D \bar{\omega}_A$ is a well-defined Lebesgue measure. In the computation of the partition function, the chosen normalization coefficient is another way to get ride of this infinite dimensional contribution. For the moment, we consider this measure as a legit Lebesgue measure even if it is well-known that in the infinite dimensional case every translation-invariant measure which is not identically zero has the property that all open sets have infinite measure . Hence, we have
	\begin{equation}
		\int DA = \sum_{n_A} \quad \sum_{\tau_A} \quad \oint d \theta_A \quad \int D \bar{\omega}_A \, . 
	\end{equation}
	We just have to performed the various integrals in order to check if \eqref{deltaonfields} is meaningful. The order in which the various operations are made should be irrelevant if decomposition \eqref{Adecomps} is consistent with integration, that is to say if the parameters of the decomposition are independent. Concerning $n_A$ and $\tau_A$ this is quite obvious. The parameter $\theta_A$ represents the quantum field $\sum_I \theta_A^I \rho_I$ with $\theta_A^I \in \mathbb{R}/\mathbb{Z}$ and $\rho_I \in \Omega^p_\mathbb{Z}$ generating $F^{p+1}$. Thus, except when trivial, this quantum field is never of the form $n_A$ or $\tau_A$. Accordingly, $\theta_A$ is independent of $n_A$ and $\tau_A$. Finally, $\bar{\omega}_A$ is by construction independent of $\theta_A$. And for the same reason as $\theta_A$, such a parameter cannot be $n_A$ or $\tau_A$. Hence, all these parameters are independent with each other, except of course if they are trivial. Let us also point out that whereas the space of quantum $p$-fields is a $\mathbb{Z}$-module, the space $\Omega^p(M)/\Omega_\circ^p(M)$ is an infinite dimensional vector space. Indeed, for any $\bar{\omega}$ and any $\lambda \in \mathbb{R}$ we set $\lambda \bar{\omega} = \overline{\lambda \omega}$. It is quite easy to check that this definition is consistent.

	\section{The Dirac delta distribution of the space of quantum fields}
	
	So, we must consider
	\begin{equation}
		\sum_{n_A} \sum_{\tau_A} \oint d \theta_A \int D \bar{\omega}_A e^{2 i \pi \oint_M (n_A \star \tau_B + n_A \star \theta_B + n_A \star \bar{\omega}_B + \tau_A \star n_B + \tau_A \star \tau_B + \theta_A \star n_B + \bar{\omega}_A \star n_B + \bar{\omega}_A \star \bar{\omega}_B)} \, .
	\end{equation}
	
	We start with the integral over $\theta_A$ which only concerns the factor $e^{2 i \pi \oint_M \theta_A \star n_B }$. It turns out that
	\begin{equation}
		\oint_M \theta_A \star n_B = \vec{\theta}_A \cdot \vec{n}_B \, ,
	\end{equation}
	where $\vec{n}_B$ represents $n_B$ in $\mathbb{Z}^{b_q} \simeq F_q(M)$ and $\vec{\theta}_B$ represents $\theta_B$ in $(\mathbb{R}/\mathbb{Z})^{b_q}$
	so that
	\begin{equation}
		\oint d \theta_A e^{2 i \pi \oint_M \theta_A \star n_B } = \delta_{\vec{n}_B,0} \, .
	\end{equation}
	Taking this into account, the original integral reads
	\begin{equation}
		\delta_{\vec{n}_B,0} \sum_{n_A} \sum_{\tau_A} \int D \bar{\omega}_A e^{2 i \pi \oint_M (n_A \star \tau_B + n_A \star \theta_B + n_A \star \bar{\omega}_B + \tau_A \star \tau_B + \bar{\omega}_A \star \bar{\omega}_B)} \, .
	\end{equation}
	
	The next obvious step is to perform the sum over $\tau_A$. Before doing this, let us recall that \cite{GT2}
	\begin{equation}
		\oint_M \tau_A \star \tau_B = - Q(\vec{\tau}_A , \vec{\tau}_B) \, ,
	\end{equation}
	where $Q$ is the non-degenerate quadratic form on $T_q(M)$ while $\vec{\tau}_A$ and $\vec{\tau}_B$ respectively represent $\tau_A$ and $\tau_B$ in $T_q(M) = \mathbb{Z}_{p_1} \oplus \mathbb{Z}_{p_2} \oplus \dots \oplus \mathbb{Z}_{p_N}$. Thus, we have
	\begin{equation}
		\sum_{\tau_A} e^{2 i \pi \oint_M \tau_A \star \tau_B} = \sum_{\vec{\tau}_A} e^{-2 i \pi Q(\vec{\tau}_A , \vec{\tau}_B)} = |T_q(M)| \delta_{\vec{\tau}_B,0} \, ,
	\end{equation}
	where $|T_q(M)|$ denotes the order of $T_q(M)$. A proof of this result can be found in \cite{MT}. With this second constraint, our original integral now reads
	\begin{equation}
		|T_q(M)| \delta_{\vec{n}_B,0} \delta_{\vec{\tau}_B,0} \sum_{n_A} \int D \bar{\omega}_A e^{2 i \pi \oint_M (n_A \star \theta_B + n_A \star \bar{\omega}_B + \bar{\omega}_A \star \bar{\omega}_B)} \, ,
	\end{equation}
	as $n_A \star \tau_B = 0$ due to the constraint $\delta_{\vec{\tau}_B,0}$.
	
	The before last step is to consider the infinite dimensional integral over $\bar{\omega}_A$. To find its expression let us first recall that
	\begin{equation}
		\oint_M \bar{\omega}_A \star \bar{\omega}_B = \oint_M \omega_A \wedge d \omega_B \, ,
	\end{equation}
	where $\omega_A$ and $\omega_B$ are representatives of $\bar{\omega}_A$ and $\bar{\omega}_B$ respectively. We now admit that such a representative has been chosen so that the integral over $\bar{\omega}_A$ is just the integral over the representives $\omega_A$. Assuming that the integration over $\omega_A$ produce the same results as a finite dimensional one, we deduce that
	\begin{equation}
		\int D \bar{\omega}_A e^{2 i \pi \oint_M ( \bar{\omega}_A \star \bar{\omega}_B)} = \int D \omega_A e^{2 i \pi \oint_M \omega_A \wedge d \omega_B} \, .
	\end{equation}
	This means that $d \omega_B = 0$ and thus that $\bar{\omega}_B = 0$. As the integral of the left-hand side of the above relation is performed over the quotient space $\Omega^p(M)/\Omega^p_\circ(M) = \Omega^p(M)/(\ker d)$, it is consistent, although formal, to write
	\begin{equation}
		\int D \bar{\omega}_A e^{2 i \pi \oint_M ( \bar{\omega}_A \star \bar{\omega}_B)} = \delta (\bar{\omega}_B) \, .
	\end{equation}
	Then, our original integral takes the form
	\begin{equation}
		|T_q(M)| \delta_{\vec{n}_B,0} \delta_{\vec{\tau}_B,0} \delta (\bar{\omega}_B) \sum_{n_A} \int D \bar{\omega}_A e^{2 i \pi \oint_M (n_A \star \theta_B + n_A \star \bar{\omega}_B)}
	\end{equation}
	
	The final step is to perform the sum over $n_A$ knowing that $\bar{\omega}_B$ is represented by a closed $p$-form. Furthermore, since $\theta_B + \bar{\omega}_B = \hat{\omega}_B$ the last factor to compute can be written as
	\begin{equation}
		\sum_{n_A} e^{2 i \pi \oint_M n_A \star \hat{\omega}_B} \, .
	\end{equation} 
	The quantity $\oint_M (n_A \star \hat{\omega}_B)$ can be determine in the following way. We consider a set of $q$-cycles, $z_I$, which generates $F_p(M)$. Then, we have
	\begin{equation}
		\oint_M n_A \star \hat{\omega}_B = \sum_I n^I_A \oint_{Z_I} \omega_B \, ,
	\end{equation}
	where the integers $n^I_A$ are the component of $\vec{n}_A$. Let us point out that since $\omega_B$ is closed, this integral is independent of $p$-cycles chosen to generate $F_p(M)$. Then, we have
	\begin{equation}
		\sum_{\vec{n}_A} e^{2 i \pi \sum_I n^I_A \oint_{Z_I} \omega_B} = \prod_I \sum_{n^I_A} e^{2 i \pi n^I_A \oint_{z_I} \omega_B} \, .
	\end{equation}
	Finally, the quantity $\oint_{z_I} \omega_B$ is by construction an element of $\mathbb{R}/\mathbb{Z}$ since $\omega_B$ represents an element of $\Omega^p(M)/\Omega_\mathbb{Z}^p(M)$. Consequently, we have
	\begin{equation}
		\sum_{\vec{n}_A} e^{2 i \pi \sum_I n^I_A \oint_{z_I} \omega_B} = \prod_I \delta (\oint_{z_I} \omega_B) \, ,
	\end{equation}
	with delta the Dirac distribution in $\mathbb{R}/\mathbb{Z} \simeq S^1$. In other words, $\oint_{z_I} \omega_B$ must be an integer for each generating cycle $z_I$. As we already saw that it must be closed, we deduce that $\omega_B \in \Omega^p_\mathbb{Z}(M)$ and thus that $\hat{\omega}_B = 0$, which also means that $\theta_B = 0$ and $\bar{\omega}_B = 0$. 
	
	At this point, it is possible to check that the order of integration is really irrelevant. Although this is an interesting exercise, this is ensured by the fact that decomposition \eqref{Adecomps} involves independent parameters, as already mentioned at the end of section 2. 
	
	If we gather all these results we conclude that our original integral can be written as
	\begin{equation}
		|T_q(M)| \, \delta_{\vec{n}_B,0} \, \delta_{\vec{\tau}_B,0} \, \delta (\theta_B) \, \delta(\bar{\omega}_B) = |T_q(M)| \delta(B) \, ,
	\end{equation}
	which up to the factor $|T_q(M)|$ is the expected result. In other words, we have
	\begin{equation}
		\delta(B) = \frac{1}{|T_q(M)|} \int \! \! D\!A \, e^{2 i \pi \oint_M A \star B} \,.
	\end{equation}
	Moreover, by construction we have
	\begin{equation}
		\int D \! B \, \delta(B) = \frac{1}{|T_q(M)|} \int \! \! D\!A \, D \! B e^{2 i \pi \oint_M A \star B} = 1 \, .
	\end{equation}
	However, we also know that
	\begin{equation}
		\mathcal{Z}_{BF_1} = |T_q(M)| \, .
	\end{equation}
	Hence, we can replace the formal normalization used in the determination of $\mathcal{Z}_{BF_1}$ by the condition
	\begin{equation}
		\int D \bar{\omega}_A e^{2 i \pi \oint_M \bar{\omega}_A \star \bar{\omega}_B} = \delta (\bar{\omega}_B) \, .
	\end{equation}
	This way to proceed seems less ad hoc than the use of a normalization which takes away this infinite dimensional integral. Nonetheless, both procedure remain formal, since applied to functional integrals.
	
	Let us now consider the case of the partition function $\mathcal{Z}_{BF_k}$ or rather of the quantum Lagrangian with quantize coupling constant $k$, i.e. $k A \star B$. On the one hand, from the previous procedure we deduce that 
	\begin{equation}
		\frac{1}{|T_q(M)|} \int \! \! D\!A \, e^{2 i \pi k \oint_M A \star B} = \delta(k B) \, ,
	\end{equation}
	since $k A \star B = A \star (k B)$. On the other hand, if we want to naively use the standard property
	\begin{equation}
		\delta(f(\vec{x})) = \sum_i \frac{\delta(\vec{x} - \vec{x}_i)}{|(\nabla f)(\vec{x}_i)|} \, ,
	\end{equation}
	with $x_i$ the roots of $f$, we first need to determine the quantum fields which are solutions of $k B =0$. Let us first remark that $\delta(k B) \neq k^{-r} \delta(B)$ for some integer $r$ because the equation $k B =0$ has more solutions than just $B = 0$.
	
	If $(n_B , \tau_B , \hat{\omega}_B)$ is the decomposition of the quantum field $B$, then the decomposition of $k B$ is obviously $(k n_B , k \tau_B , k \hat{\omega}_B)$. So, the constraint $k B =0$ implies that $k n_B = 0$ and since $n_B$ is represented by elements of $\mathbb{Z}^{b_q}$, the only possibility is $n_B = 0$. Thus, the decomposition of $B$ is of the form $(0 , \tau_B , \hat{\omega}_B)$. In the same way, we must have $k \tau_B = 0$. But since $\tau_B$ is an element of $T_{p+1}(M)$, it can be written as an $N$-tuple $(t_1 , t_2, \dots , t_N)$ of $\mathbb{Z}_{p_1} \oplus \mathbb{Z}_{p_1} \oplus \dots \oplus \mathbb{Z}_{p_N}$ so that the constraint $k \tau_B = 0$ yields the set of constraints
	\begin{equation}
		k t_I = 0 \mod p_I \, ,
	\end{equation}
	for $I \in \{1, \dots , N \}$. Quite obviously, there are $pgcd(k,p_I)$ solutions for each of the above equations. More precisely, if we set $k = pgcd(k,p_I) k_I$ and $p_I = pgcd(k,p_I) p'_I$ then the solutions of the above equation are $t_I \in \{0 , p'_I , \dots , (pgcd(k,p_I) - 1)p'_I \}$. The corresponding decompositions of $B$ are $(0 , t_B , \hat{\omega}_B)$ where $t_B$ is represented by $\vec{t}_B = (t_1 , t_2, \dots , t_N)$. The last constraint to solve is then $k \hat{\omega}_B = 0$ which means that $\hat{\omega}_B$ admits a representative $\omega_B$ such that $k \omega_B$ is closed with integral periods. This implies that $\omega_B$ is closed and thus that $\bar{\omega}_B = 0$, the very last constraint thus being $k \theta_B = 0$. The obvious solutions of this equation are
	\begin{equation}
		\vec{\epsilon}_B = \left( \frac{a_1}{k} , \frac{a_2}{k} , \dots , \frac{a_{b_q}}{k}  \right) = \frac{\vec{a}}{k} \, ,
	\end{equation}
	with $a_i \in \{0 , 1 , \dots , k-1 \}$. In conclusion, the decomposition of a quantum $p$-field $B$ such that $k B =0$ is $(0 , t_B , \epsilon_B , 0)$, with $t_B$ represented by $\vec{t}_B$ and $\epsilon_B$ by $\vec{\epsilon}_B$. Taking into account all these results, we can conclude that
	\begin{equation}
		\delta(k B) = \delta_{\vec{n}_B,0} \left(\sum_{\vec{t}_B} \delta_{\vec{\tau}_B - \vec{t}_B , 0}\right) \left(\sum_{\vec{\epsilon}_B} \delta(\vec{\theta} - \vec{\epsilon}_B)\right) \delta(\bar{\omega}_B)
	\end{equation}
	Alternatively, we can compute $\int D \bar{\omega}_A e^{2 i \pi k \oint_M \bar{\omega}_A \star \bar{\omega}_B}$ as we did in the case $k = 1$. In fact, it is quite obvious that we will obtain
	\begin{equation}
		\int D \bar{\omega}_A e^{2 i \pi k \oint_M \bar{\omega}_A \star \bar{\omega}_B} = |T_q(M)| \delta_{\vec{n}_B,0} \left(\sum_{\vec{t}_B} \delta_{\vec{\tau}_B - \vec{t}_B , 0}\right) \left(\sum_{\vec{\epsilon}_B} \delta(\vec{\theta} - \vec{\epsilon}_B)\right) \delta(\bar{\omega}_B) \, ,
	\end{equation}
	as expected. 
	
	Once more, we can relate $\delta(k B)$ with the $U(1)$ BF partition function $\mathcal{Z}_{BF_k}$ according to
	\begin{equation}
		\label{deltaexpressBF}
		\mathcal{Z}_{BF_k} = |T_q(M)| \int \! D \! B \, \delta(k B) \, .
	\end{equation}
	The right hand side of this equality uses the convention $\int D \bar{\omega}_A e^{2 i \pi \oint_M \bar{\omega}_A \star \bar{\omega}_B} = \delta (\bar{\omega}_B)$ whereas to obtain the left hand side a normalization factor was used. Let us make a final remark concerning the formal expression of $\delta (\bar{\omega}_B)$. 
	
	\section{Gauge fixing}
	
	Usually, the exponent defining the Fourier transform involves a scalar product. To write $e^{2 i \pi \oint_M \bar{\omega}_A \star \bar{\omega}_B}$ in a similar way, we can first endow $M$ with a Riemannian metric $g$ and then use this metric to write any $p$-form $\omega$ as the sum of a closed form $\omega^\circ_{(g)}$ and of a co-exact form $\omega^\perp_{(g)}$ according to
	\begin{equation}
		\label{decomptrans}
		\omega = \omega^\circ_{(g)} + \omega^\perp_{(g)} \, ,
	\end{equation}
	the closed $p$-form $\omega^\circ_{(g)}$ being itself the sum of a $g$-harmonic $p$-form and an exact $p$-form. Although they dependent on $g$, the forms $\omega^\circ_{(g)}$ and $\omega^\perp_{(g)}$ are unique for a given metric $g$. Decomposition \eqref{decomptrans} implies that any element of $\Omega^p(M)/\Omega_\circ^p(M)$ can be represented by a co-exact $p$-form, and we have
	\begin{equation}
		\oint_M \bar{\omega}_A \star \bar{\omega}_B = \oint_M \omega^\perp_A \wedge d \omega^\perp_B \, ,
	\end{equation}
	where the reference to $g$ was removed to lighten the notations. The right hand side of the above identity can be written as a scalar product according to
	\begin{equation}
		\oint_M \omega^\perp_A \wedge d \omega^\perp_B = <\omega^\perp_A , ^\ast d \omega^\perp_B > \, ,
	\end{equation}
	where $\ast$ is the Hodge operation associated with the metric $g$. Finally, we have
	\begin{equation}
		\label{deltaHodege}
		\int D \bar{\omega}_A e^{2 i \pi k \oint_M \bar{\omega}_A \star \bar{\omega}_B} = \int D \omega^\perp_A e^{2 i \pi k <\omega^\perp_A , ^\ast d \omega^\perp_B >} \, ,
	\end{equation}
	where the functional integral in the right-hand side of the above identity is performed over the space of co-exact $p$-forms $\omega^\perp_A$. Its ``value" is $\delta (d \omega^\perp_B)$. This means that beside being co-exact by construction, the $p$-form $\omega^\perp_B$ is closed. This implies that $\omega^\perp_B = 0$ and hence that $\bar{\omega}_B = 0$, as expected. The interesting point of this approach which yields the factor $\delta(\bar{\omega}_B)$ is that it naturally gives rise to representatives of elements of $\Omega^p(M)/\Omega_\circ^p(M)$. The drawback is that these representatives depend on the metric even if the final result does not. This reminds us of the gauge fixing procedure used in QFT. Let us note that $\Omega^p(M)/\Omega_\circ^p(M)$ is a (topological) vector space and not just a $\mathbb{Z}$-module. Hence, it is quite natural to find that the right-hand side of \eqref{deltaHodege} is $\delta (d \omega^\perp_B)$. Moreover, the right-hand side of the above identity must not be confused with the functional integral $\int D \omega_A e^{2 i \pi k <\omega_A , ^\ast d \omega_B >}$. Such an integral is also ill-defined not just because it is infinite dimensional but also because the quantity $<\omega_A , ^\ast d \omega_B > = \oint_M \omega_A \wedge d \omega_B$ is gauge invariant, the gauge group being precisely $\Omega_\circ^p(M)$, which is also infinite dimensional. The  ``volume" of $\Omega_\circ^p(M)$ must be factorized out of the functional integral. In the particular case where $\Omega_\circ^p(M) = d \Omega^{p-1}(M)$, a Faddeev-Popov procedure can be applied \cite{M19}. In the case where $F^p(M)$ is not trivial, we refer to \cite{S79} for a treatment of the functional integral $\int D \omega_A e^{2 i \pi k <\omega_A , ^\ast d \omega_B >}$.
	
	Now that we have identify the formal expression of the Dirac delta distribution on the space of quantum fields of $M$, we can use it to relate the BF and Chern-Simons partition functions by writing
	\begin{equation}
		\int D \! A \, e^{2 i \pi k \oint_M A \star A} = \iint \! \! D\!A \,  D\!B \, \delta(B - A) \, e^{2 i \pi k \oint_M A \star B} \, .
	\end{equation}
	This yields
	\begin{equation}
		\label{backtoCS}
		\int D \! A \, e^{2 i \pi k \oint_M A \star A} = \iint \! \! D\!A \,  D\!B \, DC \, e^{2 i \pi k \oint_M (C \star B + A \star B - C \star A)} \, .
	\end{equation}
	It is a tedious but simple exercise to prove that the above relation is true by using the standard decomposition of the various quantum fields appearing in the left and right hand sides and then performing the appropriate integrals.

	Expressing the BF partition function as in \eqref{deltaexpressBF} is also another way to see the relation between this partition function and the $\mathbb{Z}_k$ TV invariant of $M$. Indeed, this last invariant of $M$ is a product of Kronecker delta symbols \cite{MT}. Furthermore, this same invariant can also be written as the partition function of a discrete BF theory \cite{MT2}. We also know that it is possible to obtain the $\mathbb{Z}_k$ TV invariant via a Rechetikhin-Turaev procedure which is based on surgery \cite{RT} whereas the TV construction is based on a cellular decomposition of $M$ \cite{TV,BK}.

	\section{Conclusion}
	
	In this article we have shown how a Dirac delta distribution on the space of quantum fields of the $U(1)$ BF theory can be introduced as a functional integral on this space, in a way which recall the link between Fourier transform and the usual Dirac distribution. The BF and Chern-Simons partition functions have then been revisited with the help of this Dirac delta distribution. As a consequence, we get rid of the normalisation trick usually used in determining these partition functions. The fact that the Chern-Simons partition function can be expressed with some BF actions only suggests that it could be possible to obtain the abelian RT invariant from a cellular decomposition of $M$ instead of a surgery procedure. However, in order to write the TV invariant as a discrete partition function we need, beside a cellular decomposition of $M$, a dual decomposition. This implies that the quantities which correspond to the quantum fields $A$ and $B$ do not belong to the same decomposition and hence it is not obvious to write the equivalent of $A \star A$ in this context.
	
	As a final remarks, let us recall that the the TV invariant of a closed $3$-manifold $M$ is defined with the help of the so called total state spaces of a polyhedral decomposition of $M$, each total state space being indexed by a labeling of the decomposition. In the abelian case, total state spaces are simply defined by products of delta symbols, the TV invariant being a sum over the labelings of these products \cite{MT}. Thus, by choosing to write the $U(1)$ BF partition function as a functional integral over quantum fields of Dirac delta -- as in \eqref{deltaexpressBF} -- its relation with the TV construction is clearly highlighted.

	\vfill\eject

\end{document}